# Y-net: 3D intracranial artery segmentation using a convolutional autoencoder

Li Chen[1], Yanjun Xie[2], Jie Sun[3], Niranjan Balu[3], Mahmud Mossa-Basha[3], Kristi Pimentel[3], Thomas S. Hatsukami[4], Jenq-Neng Hwang[1], Chun Yuan[3†]

Department of 1. Electrical Engineering; 2. Mechanical Engineering; 3. Radiology; 4. Surgery
University of Washington
Seattle, WA, 98195, USA
{cluw, yanjunx. sunjie, ninja, mmossab, kristidb, tomhat, hwang, cyuan }@uw.edu
†Corresponding author

*Abstract*— **Automated segmentation of intracranial arteries on magnetic resonance angiography (MRA) allows for quantification of cerebrovascular features, which provides tools for understanding aging and pathophysiological adaptations of the cerebrovascular system. Using a convolutional autoencoder (CAE) for segmentation is promising as it takes advantage of the autoencoder structure in effective noise reduction and feature extraction by representing high dimensional information with low dimensional latent variables. In this report, an optimized CAE model (Y-net) was trained to learn a 3D segmentation model of intracranial arteries from 49 cases of MRA data. The trained model was shown to perform better than the three traditional segmentation methods in both binary classification and visual evaluation.**

## I. INTRODUCTION

A healthy cerebrovascular system that delivers sufficient blood flow to all brain territories is of vital importance for maintaining brain health. Indeed, the pathogeneses of many common neurological diseases, such as stroke and dementia, have been revealed to involve significant vascular contributions [1]. Free from contrast medium and ionizing radiation, 3D time-of-flight (TOF) magnetic resonance angiography (MRA) is the most widely-used neurovascular imaging technique, which is able to display a comprehensive map of the cerebrovascular tree. Automated segmentation of the cerebrovascular tree on TOF MRA allows for quantification of global and territorial vascular features, which can be applied to serial images to understand changes in the cerebrovascular system under various physiological and pathological conditions.

However, it is challenging to accurately segment intracranial arteries on TOF MRA given the complex network of cerebral arteries with substantial inter-individual variations [2], and weak signals in small arteries due to slow or in-plane blood flow.

Previous automated intracranial artery segmentation and tracing methods have mostly relied on pattern recognition or model-based approaches such as vesselness filters [3], region growing [4] and fast marching [5]. Some of these methods were shown to be effective in a limited anatomical region or for a specific patient group. However, as traditional image processing methods depend heavily on human specified vessel descriptors, it remains difficult to generate a comprehensive cerebrovascular map with consistently high performance for general cases. A more robust method is needed for intracranial artery segmentation.

Recently, neural networks have shown advantages over traditional medical image segmentation methods, such as in skull stripping [6], and brain tumor segmentation [7]. These neural-network-based methods allow computational models composed of multiple processing layers to learn representations of data with multiple levels of abstraction and discover intricate structures in large data sets.

An auto-encoder (AE) model is a neural network used for unsupervised learning of efficient coding. The encoder part transforms an input into a typically lower-dimensional representation, and the decoder part is tuned to reconstruct the initial input from this representation [8]. The purpose of an AE was to learn a representation from a set of data, instead of predicting the target given input. However, Vincent et al. modified the AE to a denoising autoencoder (DAE) [9] to restore artificially corrupted input, as the model is trained to extract more useful features from the original input to solve the much harder denoising problem. For this application, the model is trained in a supervised way to predict the target rather than learning representations. When used in image processing, a convolutional autoencoder (CAE), one of the DAE models using convolutional layers and deconvolution layers, is an efficient and promising structure, as it fully utilizes the advantages of convolutional neural networks (CNNs), which have proven to be effective with noisy, shifted or corrupted image data [10], allowing patterns to be learned from local pixels and positioned at various locations.

In medical image processing field, successful applications of CAE were recently reported on cell and neuronal structure segmentations [11]. By combining localization information and augmenting with elastic deformations, the authors proved their model achieved better performance than sliding-window CNN [12]. However, the intracranial artery structure is complicated and segmentation with fine details is needed. Thus, elastic deformations are unsuitable. In addition, their model is designed for 2D images.

In this report, an optimized CAE structure (Y-net) designed for 3D patch based CAE training and prediction is illustrated. Y-net is expected to denoise and extract useful features from the 3D patches extracted from large original images, and thus to perform the intracranial artery segmentation on 3D TOF MRA.

## II. METHODS

### A. Data

Forty-nine brain TOF MRA images (620*620*243 in size, resolution of 0.3 mm$^3$ per voxel) were acquired on a 3T Philips (Philips Healthcare, Best, the Netherlands) Ingenia MRI scanner using a standard head coil. Imaging parameters were: TR/TE = 14.7/3.5ms, flip angle = 18°. The study was performed with IRB approval and informed consents were obtained for all participating subjects.

### B. Preprocessing

We applied intensity correction [13] to correct image inhomogeneity and intensity normalization [14] to unify intensity ranges of all subjects between 0 to 1.

All the images were processed semi-automatedly [15]. Arteries were first traced by a deformable model. Then, vascular connections were manually examined to finalize the artery tracing results (3D positions along the centerline with radius). The traces were further verified by an experienced radiologist. 3D labeled images were then created as ground truth by filling vessel surfaces rendered in the Visualization ToolKit (VTK) [16].

A representative case of the generated labeled image patches, along with the original image patches, is shown in Fig. 1.

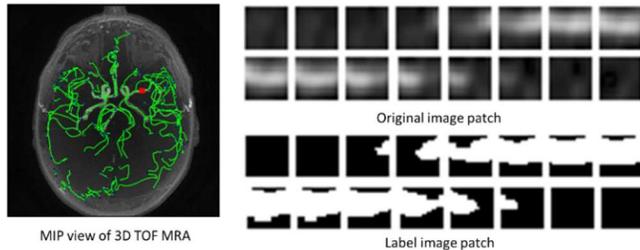

Fig. 1. A representative case of the original image patches (top right) and the corresponding labeled image patches (bottom right) shown in 2D slices. The origin of the patch is shown in a red square in the Maximum Intensity Projection (MIP) view of TOF MRA with arteries traced in green (left).

### C. Data separation

Forty-nine pairs of original and corresponding labeled images were randomly separated into three individual sets for unbiased validation and evaluation: 46 pairs for the training set, 2 pairs for the validation set (tuning network parameters), 1 pair for the test set (evaluating performance).

### D. Network structure

With 3D TOF MRA images as input and binary coded 3D labeled images as output, our task can be categorized as a supervised learning for a two-class classification task.

Instead of using the whole 3D images as input and output, we used a cubic window of 16*16*16 voxels to slide through original and labeled images to extract a batch of image patches for training and prediction. The reasons for using patches instead of the whole 3D images are as follows: 1) Intracranial arteries have similar tubular structures; 2) Patches extraction significantly increases training samples and reduces the neural network scale; 3) The diameter of largest arteries rarely exceeds 16 pixels; 4) Human can usually determine artery regions within this patch size; 5) Computational efficiency and memory saving are considered.

To train a CAE model according to our patch based appilcation, Y-net was designed from the basic structure of the original CAE, as shown in Fig. 2.

In the encoding path, there are two consecutive blocks of convolution layers (extracting useful features along with the increased number of kernels) followed by max-pooling layers (reducing dimension to remain most significant features).

In the decoder path, there are the same number of consecutive blocks of convolution layers followed by up-sampling layers to restore the dimension and combine convolution kernel responses. With similar ideas in [11], in order to localize, the corresponding feature map from the encoding path is concatenated and convolved with the up-sampled layer before the successive convolution layer assembles a more precise output based on this information.

The position path provides additional information of patch origin. The center 3D position of the patch extracted from the original image is normalized between 0 to 1 and repeated to the same size of the last encoded layer as position kernels, which are then concatenated with output in encoding path.

With the encoding and decoding paths more or less symmetric to each other, and an additional position path concatenated in the middle, the network looks like a Y shape.

Convolution kernels are cubes of 3. Max-pooling and up-sampling kernels are cubes of 2. Each convolution layer is followed by a rectified linear unit (ReLU) as the activation function. All the network parameters are initialized using Xavier initialization [17].

### E. Training process

The patch extraction method is a critical step to train this patch based CAE. If consecutive sliding is applied in each direction one voxel per sliding, even one original image can yield more than 80 million patches. However, largely unbalanced labeled patches (most patches contain no artery voxels) causes bias in training, and highly similar neighboring patches leads to longer training time. On the other side, a large stride may miss many regions with artery voxels and cause insufficient training samples.

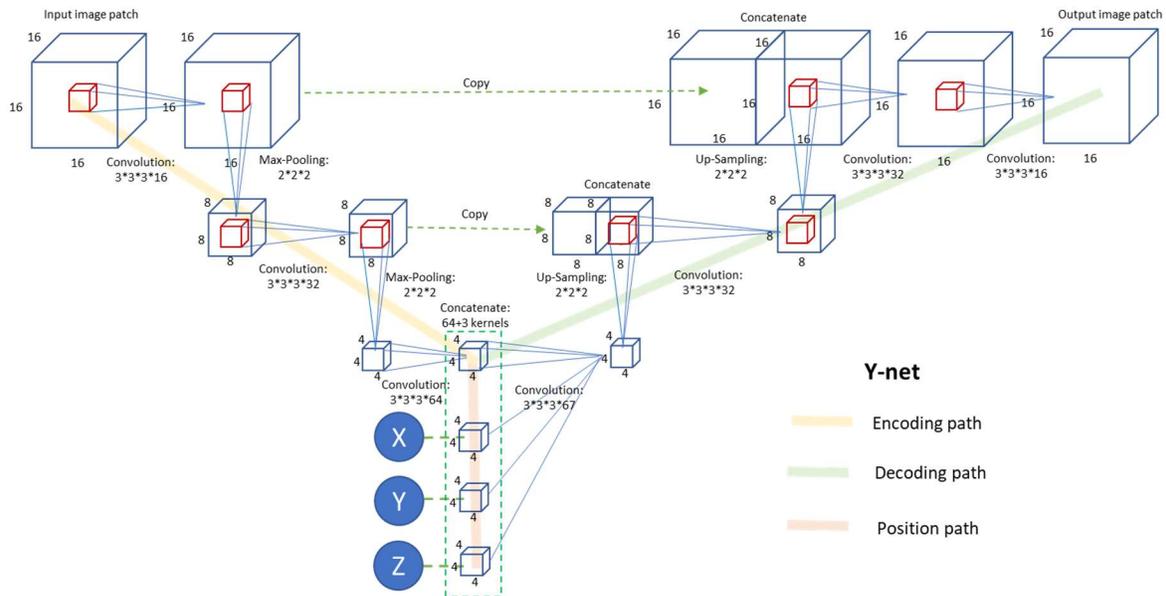

Fig. 2. Structure of the proposed Y-net CAE model.

A varied sliding window is used to extract patches. A smaller stride first extracts patches if there is at least one voxel labeled as vascular region in labeled patches; then a larger stride is estimated according to the selection rate in the previous step to extract similar number of all non-artery patches. As the number of patches trained in one batch is restricted by GPU memory, stride size in first step can be adjusted accordingly to ensure patch number extracted from most images are below the batch size.

We define one epoch as patches are extracted from each of the images in training set once. Images from training set are shuffled in each epoch.

To fully use the information between sliding windows, an offset starting from zero to stride size is used for each epoch during patch extraction.

Training period is set for 30*(stride size) epochs. The model was checked on validation set every 5 epochs. The models with improved validation loss were saved.

A parallel patch extraction and network training method was used for more efficient training.

Binary cross-entropy is used as loss function. Adam optimizer [7] is used to control the learning rate.

### F. Prediction process

Test images were cut into patches with the same sizes as the training data with half size overlapping in any direction, as input to the CAE. The prediction of CAE is a probability image showing likelihood of voxels to be artery regions. The predicted values from overlapped voxels were averaged when combining patches back to the original image size.

We applied a threshold to the probability image to generate binary prediction results for comparison with the labeled image. The threshold value giving the highest accuracy rate from the validation sets was applied for prediction in the test image.

### G. Parameter optimization

Several different scenarios are tried for optimizing a list of parameters (or choices) for our CAE model.

The training parameters we optimized are: 1) activation function: ReLU, hyperbolic tangent (tanh), or the sigmoid function; 2) dimension change times; 3) kernel number of first and last convolution layers; 4) down-sampling method: max-pooling followed by convolution, or convolution with stride; 5) up-sampling method: padding with neighbor voxels followed by convolution, or deconvolution with stride; 6) number of images used for training; 7) concatenation location of the position path: first layer of the encoding path, last layer of the encoding path, or last layer of the decoding path; 8) parameter initialization method: Xavier, or random with normal distribution; 9) post-prediction morphological operations: none, closing operation (dilating, then eroding), or opening operation (dilating, then eroding).

### H. Performance evaluation

The post-processed binary image was compared with the labeled image voxel by voxel. Binary classification performance was evaluated by accuracy, sensitivity, specificity, precision, and the dice similarity co-efficient (DSC), defined as

$$DSC = \frac{2(A \cap B)}{(A + B)}$$

where A is the ground truth result and B is the segmentation result. DSC ranges from 0 (no overlap) to 1 (identical results). DSC > 0.7 indicates excellent agreement [18].

As a reference, we also evaluated segmentation results by three traditional image processing methods: 1) Renyi Entropy method [19], the best among 16 threshold methods in the ImageJ toolkit [20]; 2) Phansalkar local threshold [21], the best among 9 local threshold methods in ImageJ; 3) the Frangi vesselness filter [3].

## I. Computational setup

Codes were written using Tensorflow [22] and Keras [23]. The optimized CAE model was trained on a workstation with Intel Xeon (Intel, Santa Clara, California, USA) E5-2630 v3 @2.4GHz 8 cores, 32 GB Memory, NVIDIA GeForce (Nvidia Corporation, Santa Clara, California, USA) GTX TITAN X on Windows 7 (Microsoft Corporation, Redmond, Washington, USA).

## III. EXPERIMENTAL RESULT

### A. Parameter optimization

Training parameters were tuned and metrics were calculated for each option. The final optimized choices are: activation with ReLU, down/up sampling blocks of 2, kernel numbers of 16, down-sampling with max-pooling, up-sampling using neighbor padding before convolution, training with all images in training set, position path concatenated to last layer of encoding path, and no morphological operation.

### B. Comparison with traditional methods

The binary classification results from CAE (with variations) as well as traditional segmentation methods are shown in TABLE I.

TABLE I  BINARY CLASSIFICATION EVALUATION BETWEEN CAE (WITH THRESHOLD) AND CLASSICAL SEGMENTATION METHODS

| Models | Accuracy | Sensitivity | Specificity | Precision | DSC |
|---|---|---|---|---|---|
| CAE (0.38) | 0.99837 | 0.83815 | 0.99913 | 0.81883 | 0.82838 |
| CAE (0.15) | 0.99726 | 0.92235 | 0.99761 | 0.64560 | 0.75955 |
| CAE (0.38) no localization | 0.99823 | 0.82208 | 0.99906 | 0.80465 | 0.81327 |
| CAE (0.38) epoch 100 | 0.99801 | 0.80282 | 0.99893 | 0.77965 | 0.79107 |
| Renyi | 0.99500 | 0.52331 | 0.99724 | 0.47272 | 0.49673 |
| Phansalkar | 0.85897 | 0.92017 | 0.85868 | 0.02992 | 0.05795 |
| Frangi | 0.99731 | 0.56023 | 0.99937 | 0.80856 | 0.66187 |

### C. Visual evaluation of segmented results

A threshold of 0.38 was selected. The generated MIP views of the labeled image, predicted image by CAE, and segmentation results using traditional methods of the test set in VTK as shown in Fig. 3.

The posterior communicating artery (PComm), a small artery that is easily missed but clinically important, is examined by observing segmentation results as shown in Fig. 4. CAE correctly identified PComm when it was missed by the Renyi method. Frangi filtered artery had the discontinuity problem near bifurcations.

### D. Processing time

The batch size was found to be limited to 2048 for our hardware, and the stride size in first step of patch extraction was set to 20. The total number of patches extracted in one epoch is about 70,000.

The training time was 745 minutes. Prediction for each image took about one minute.

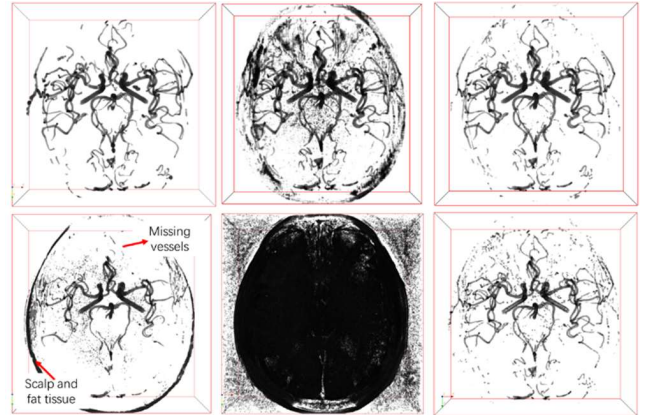

Fig. 3. Top Left: Labeled image. Top middle: Predicted image (with possibility from 0 to 1) by CAE. Top Right: Predicted image after thresholding. Bottom Left to Right: Segmentation results by Renyi Entropy threshold, Phansalkar local threshold, and Frangi filtering.

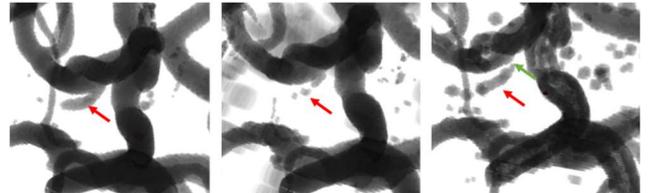

Fig. 4. Segmentation results near the left PComm (pointed by red arrows) predicted by CAE (Left), Renyi Entropy Threshold (Middle), and Frangi Filter (Right). Green arrow points at the position where PComm discontinues at bifurcation. Phansalkar method is too noisy for visualization.

## IV. DISCUSSION

### A. Network structure

AE models have typically been used as an unsupervised learning approach to extract useful features. For our supervised learning application, the idea of extracting useful features from voxels is similar, but we are more interested in the performance of the final decoded results.

Compared with the traditional and previous CAE models, our Y-net CAE has several major optimizations which allows effective and efficient intracranial artery segmentation.

1) We use patch based images to train the network, considering the anatomical prior from intracranial arteries. The benefits are two folded. From the perspective of training data, extracting patches from large images can easily avoid the usual problem of insufficient training samples. Network can converge even without risky data augmentation. Different from cell images which can be elastically deformed, intracranial arteries are a complex vascular network with twist and twine, thus unsuitable for deformed augmentation. From the perspective of network structure, smaller input and output reduce the network size and parameter number, as a result, no pre-trained models are needed, and the model can be learned from scratch in a reasonable time.

2) Position path is concatenated into the CAE structure to ensure localization information remains after patch extraction. Some extracranial arteries or veins have very similar anatomical structures as intracranial arteries if focusing on patch region alone. However, these false positive targets might be eliminated if patch origin is included.

3) We used 3D patches and operations in our solution. Vascular segmentation is different from tumor or skull segmentation in that the target is relatively small in volume and complex in shape. To segment fine details, a 3D neural network model is necessary. In some 3D applications [6], 2D slices of 3D images were trained, predicted, then stacked together. Despite increased calculation time, additional information between slices contributed to consistent segmentation between slices.

*B. Training process*

Accurately labeled images are critical in training a CAE models. However, labels are difficult to acquire for intracranial artery segmentation. It is time-consuming for humans to manually segment 3D artery regions, and it is equally challenging for automated algorithms to consistently generate high quality artery segmentation results. To balance time efficiency and segmentation quality, we used a semi-automated method to trace arteries in 3D space with human intervention and correction as the ground truth. It still takes more than one hour to trace and validate one case.

We found the training loss and validation loss decrease mainly in the first 100 epochs, by which time the performance already exceeds traditional methods. Our chosen training period (600 epochs) is long enough for the model to be stable.

*C. Comparison with traditional methods*

As accuracy alone is not enough to evaluate the model performance when 99.4% of voxels are non-artery regions, five metrics were used for comparison between models. We found CAE outperformed the Renyi Entropy and Frangi filter methods in all the five metrics. While the Phansalkar method has high sensitivity but low precision. If the threshold is set to 0.15, CAE outperforms Phansalkar method in all metrics.

From visual evaluation of the segmentation results, we found the CAE method removed most scalp and fat tissue around the brain while maintained the main structure of artery tree and even small arteries, such as the PComm. Still, some arteries were not segmented continuously, especially for peripheral segments of middle cerebral arteries where signal intensity becomes relatively low. Nonetheless, the overall visualization effect of the CAE method is superior to the traditional methods.

Compared with traditional methods, another benefit of the neural network structure used in CAE is its adaptiveness to data. With little need for optimization, this model is also applicable to other images and has potential for segmentation tasks in other modalities.

*D. Future work*

While CAE showed excellent agreement with ground truth (DSC: 0.83) in intracranial artery segmentation, additional improvement is possible. Future improvements may target larger dataset, better labeling methods, and improved network models.

## V. CONCLUSION

We trained a convolutional autoencoder for intracranial artery segmentation on brain TOF MRA. Our Y-net CAE model outperformed three traditional segmentation methods in both binary classification and visual evaluation. Accurate segmentation of the whole cerebral vasculature will facilitate quantification of global and territorial vascular features.

ACKNOWLEDGMENT

We appreciate the following funding agencies or companies for supporting this project: National Institutes of Health, the American Heart Association, and the Philips. We are grateful to the support of NVIDIA Corporation with donation of the Titan X GPU.